\documentclass[preprint,showpacs,preprintnumbers,amsmath,amssymb]{revtex4}
\usepackage{amssymb}
\usepackage{amsmath}
\usepackage{graphicx}
\usepackage{epsfig}
\usepackage{subfigure}
\usepackage{amsfonts}
\usepackage{CJK}

\begin{document}

\title{Circuit QED: single-step realization of a multiqubit controlled phase gate with one microwave photonic qubit simultaneously
controlling $n-1$ microwave photonic qubits}

\author{Biaoliang Ye$^{1}$}
\author{Zhen-Fei Zheng$^{2}$}
\author{Yu Zhang$^{3}$}
\author{Chui-Ping Yang$^{1,4}$}
\email{yangcp@hznu.edu.cn}

\address{$^1$Quantum Information Research Center, Shangrao Normal University, Shangrao 334001, China}
\address{$^2$CAS Key Laboratory of Quantum Information, University of Science and Technology of China, Hefei 230026, China}
\address{$^3$School of Physics, Nanjing University, Nanjing 210093, China}
\address{$^4$Department of Physics, Hangzhou Normal University, Hangzhou 310036, China}

\begin{abstract}
We present a novel method to realize a multi-target-qubit controlled phase gate with
one microwave photonic qubit simultaneously controlling $n-1$ target
microwave photonic qubits. This gate is implemented with $n$ microwave cavities coupled to a
superconducting flux qutrit. Each cavity hosts a microwave photonic qubit, whose two
logic states are represented by the vacuum state and the single photon state of a single cavity mode, respectively.
During the gate operation, the qutrit remains in the ground
state and thus decoherence from the qutrit is greatly suppressed. This proposal requires
only a single-step operation and thus the gate implementation is quite simple. The gate operation time is
independent of the number of the qubits. In addition, this proposal does not need applying classical pulse or any measurement. Numerical
simulations demonstrate that high-fidelity realization of a controlled phase gate with
one microwave photonic qubit simultaneously controlling two target microwave photonic qubits is
feasible with current circuit QED technology. The proposal is quite general and can be applied to implement
the proposed gate in a wide range of physical systems, such as multiple microwave or optical
cavities coupled to a natural or artificial $\Lambda$-type three-level atom.
\end{abstract}
\maketitle
\date{\today}

\section{Introduction}
Multiple qubit gates play important roles and are a crucial element in
quantum information processing (QIP). A multiqubit gate can in principle be
decomposed into a series of two-qubit and single-qubit gates, and thus can
be constructed by using these basic gates. However, it is commonly
recognized that building a multiqubit gate is difficult via the conventional
gate-decomposition protocol. This is because the number of basic gates,
required for constructing a multiqubit gate, increases drastically as the
number of qubits increases. As a result, the gate operation time would be
quite long and thus the gate fidelity would be significantly decreased by
decoherence. Hence, it is worthwhile to seek efficient approaches to realize
multiqubit quantum gates. Many efficient schemes have been presented for the
direct realization of a multiqubit controlled-phase or controlled-NOT gate,
with multiple-control qubits acting on one target qubit [1--14]. This type
of multiqubit gate is of significance in QIP, such as quantum algorithms and
error corrections.

In this work, we focus on another type of multiqubit gate, i.e., a
multi-target-qubit controlled phase gate with one qubit simultaneously
controlling multiple target qubits. This multi-target-qubit controlled phase
gate is described by
\begin{eqnarray}
\left\vert 0_{1}\right\rangle \left\vert i_{2}\right\rangle \left\vert
i_{3}\right\rangle ...\left\vert i_{n}\right\rangle &\rightarrow &\left\vert
0_{1}\right\rangle \left\vert i_{2}\right\rangle \left\vert
i_{3}\right\rangle ...\left\vert i_{n}\right\rangle ,  \notag \\
\left\vert 1_{1}\right\rangle \left\vert i_{2}\right\rangle \left\vert
i_{3}\right\rangle ...\left\vert i_{n}\right\rangle &\rightarrow &\left\vert
1_{1}\right\rangle \left( -1\right) ^{i_{2}}\left( -1\right)
^{i_{3}}...\left( -1\right) ^{i_{n}}\left\vert i_{2}\right\rangle \left\vert
i_{3}\right\rangle ...\left\vert i_{n}\right\rangle ,
\end{eqnarray}%
where $i_{2},i_{3},...i_{n}\in \left\{ 0,1\right\} ;$ subscript $1$
represents the control qubit while subscripts $2,3,\ldots ,$ and $n$
represent target qubits. From Eq. (1), it can be seen that when the control
qubit $1$ is in $\left\vert 1\right\rangle $, a phase flip (from sign $+$ to
$-$) happens to the state $\left\vert 1\right\rangle $ of each of target
qubits $2,3,\ldots ,$ and $n;$ however nothing happens to the states of each
of target qubits $2,3,\ldots ,$ and $n$ when the control qubit $1$ is in the
state $\left\vert 0\right\rangle $.

This multiqubit gate (1) is useful in QIP, such as entanglement preparation
[15], error correction [16], quantum algorithms [17], and quantum cloning
[18]. How to efficiently implement this multiqubit gate becomes necessary
and important. Over the past years, based on cavity QED or circuit QED, many
efficient methods have been proposed for the direct implementation of this
multiqubit phase gate, by using natural atoms or artificial atoms (e.g.,
superconducting qubits, quantum dots, or nitrogen-vacancy center ensembles)
[19--23].

Circuit QED is analogue of cavity QED, which consists of superconducting
qubits and microwave resonators or cavities. It has developed fast
recently and is considered as one of the most promising candidates for QIP
[23--29]. Owing to the microfabrication technology scalability, individual
qubit addressability, and ever-increasing qubit coherence time [30--38], superconducting
qubits are of great importance in QIP. The strong and ultrastrong couplings
between a superconducting qubit and a microwave cavity have been experimentally
demonstrated [39,40]. For a review on the ultrastrong coupling, refer to [41]. On the other hand, a (loaded) quality factor $Q\sim
10^{6}$ has been experimentally reported for a one-dimensional coplanar
waveguide microwave resonator [42,43], and a (loaded) quality factor $Q\sim
3.5\times 10^{7}$ has also been experimentally reported for a
three-dimensional microwave cavity [44]. A microwave resonator or cavity
with the experimentally-reported high quality factor here can act as a good
quantum data bus [45--47] and be used as a good quantum memory [48,49],
because it contains microwave photons whose lifetimes are much longer than
that of a superconducting qubit [50]. These good features make microwave resonators or
cavities as a powerful platform for quantum computation and microwave
photons as one of promising qubits for QIP. Recently, quantum state
engineering and QIP with microwave fields or photons have become
considerably interesting [51-72].

Motivated by the above, we will propose a method to realize the
multi-target-qubit controlled phase gate (1) with microwave photonic qubits,
by using $n$ microwave cavities coupled to a superconducting flux qutrit (a $%
\Lambda $-type three-level artificial atom) (Fig.~1). Note that to simplify the
presentation, we will use \textquotedblleft MP qubit" to denote \textquotedblleft microwave photonic qubit" and
\textquotedblleft MP qubits" to define \textquotedblleft microwave photonic qubits".
This work is based on circuit QED. As shown below, this proposal has the following advantages: (i)
During the gate operation, the qutrit stays in the ground state and thus
decoherence from the qutrit is greatly suppressed; (ii) Because of only
using one-step operation , the gate implementation is quite simple; (iii) Neither
classical pulse nor measurement is required; (iv) The gate operation time is
independent of the number of the qubits; and (v) This proposal is quite general and can be
extended to a wide range of physical systems to realize the proposed gate,
such as multiple microwave or optical cavities coupled to a natural or
artificial $\Lambda $-type three-level atom. To the best of our knowledge,
this work is the first to show the one-step implementation of a multi-target-qubit
controlled phase gate with MP qubits, based on cavity- or
circuit-QED and without any measurement.

This paper is organized as follows. In Sec. II, we explicitly show how to
realize a controlled-phase gate with one MP qubit
simultaneously controlling $n-1$ target MP qubits. In Sec.
III, we discuss the experimental feasibility for implementing a three-qubit
controlled phase gate, by considering a setup of three one-dimensional
transmission line resonators coupled to a superconducting flux qutrit. We end up with a
conclusion in Sec. IV.

\begin{figure}[tbp]
\begin{center}
\includegraphics[bb=65 400 534 590, width=12.5 cm, clip]{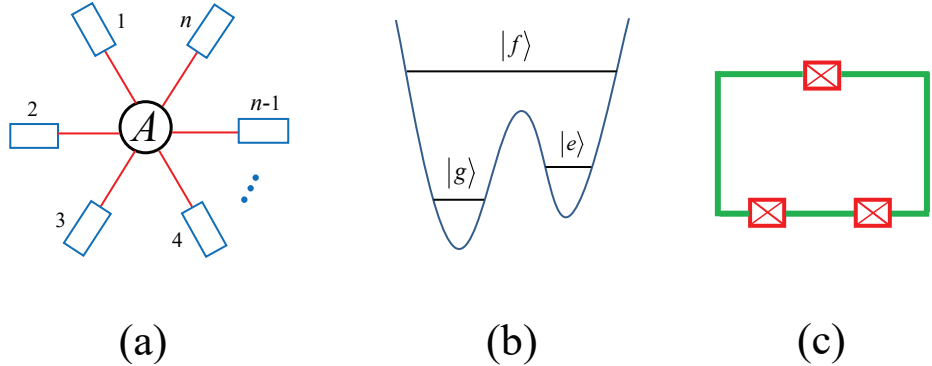} \vspace*{%
-0.08in}
\end{center}
\caption{(a) Diagram of $n$ cavities ($1,2,...,n$) coupled to
a superconducting flux qutrit \textit{A}. A square represents a cavity,
which can be a one-dimensional or three-dimensional cavity. The qutrit is
capacitively or inductively coupled to each cavity. (b) Level configuration
of the flux qutrit, for which the transition between the two lowest levels
can be made weak by increasing the barrier between two potential wells. (c)
Diagram of a flux qutrit, which consists of three Josephson junctions and a
superconducting loop.}
\label{fig:1}
\end{figure}

\section{Multi-Target-Qubit controlled phase gate}

Consider $n$ microwave cavities ($1,2,...,n$) coupled to a superconducting
flux qutrit [Fig.~1(a)]. The three levels of the qutrit are denoted as $%
|g\rangle $, $|e\rangle $ and $|f\rangle $, as shown in Fig.~1(b). In
general, there exists the transition between the two lowest levels $%
|g\rangle $ and $|e\rangle $, which, however, can be made to be weak by
increasing the barrier between the two potential wells. Suppose that cavity $%
1$ is dispersively coupled to the $|g\rangle \leftrightarrow |f\rangle $
transition of the qutrit with coupling constant $g_{1}$ and detuning $\delta
_{1}$ but highly detuned (decoupled) from the $|e\rangle \leftrightarrow
|f\rangle $ transition of the qutrit. In addition, assume that cavity $l$ ($%
l=2,3,...,n$) is dispersively coupled to the $|e\rangle \leftrightarrow
|f\rangle $ transition of the qutrit with coupling constant $g_{l}$ and
detuning $\delta _{l}$ but highly detuned (decoupled) from the $|g\rangle
\leftrightarrow |f\rangle $ transition of the qutrit (Fig.~2). Note that
these conditions can be satisfied by prior adjustment of the qutrit's level
spacings or/and the cavity frequency. For a superconducting qutrit, the
level spacings can be rapidly (within 1-3 ns) tuned by varying external
control parameters (e.g., magnetic flux applied to the loop
of a superconducting phase, transmon [73], Xmon [33], or flux qubit/qutrit [74]). In
addition, the frequency of a microwave cavity or resonator can be rapidly
adjusted with a few nanoseconds [75,76].

\begin{figure}[tbp]
\begin{center}
\includegraphics[bb=181 479 417 736, width=8.0 cm, clip]{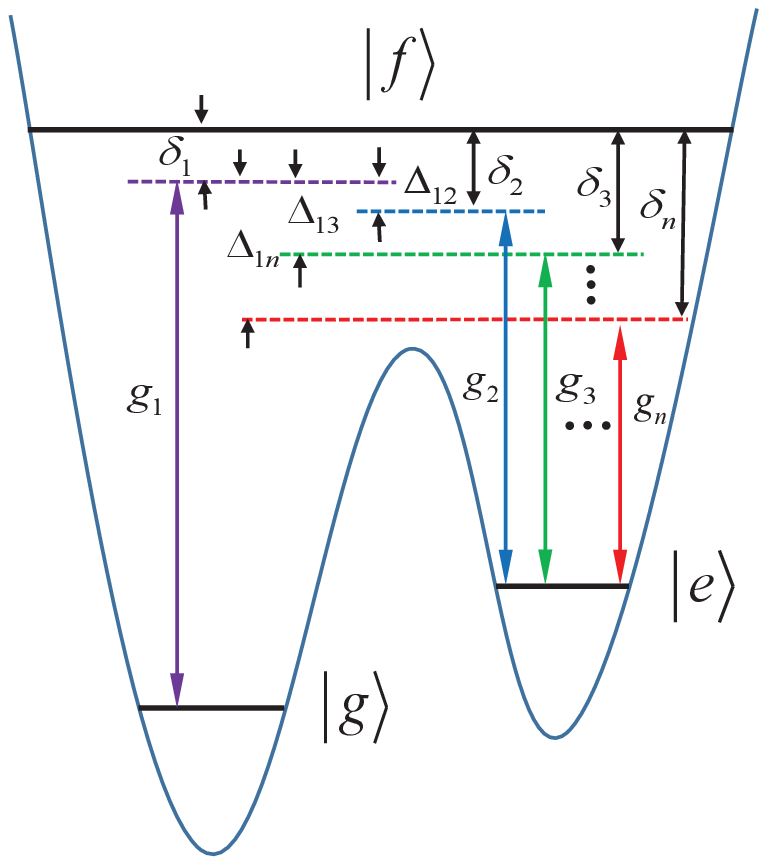} \vspace*{%
-0.08in}
\end{center}
\caption{Cavity $1$ is dispersively coupled to the $|g\rangle
\leftrightarrow |f\rangle $ transition of the qutrit with coupling strength $%
g_{1}$ and detuning $\protect\delta _{1}$, while cavity $l$ ($l=2,3,...,n$)
is dispersively coupled to the $|e\rangle \leftrightarrow |f\rangle $
transition of the qutrit with coupling strength $g_{l}$ and detuning $%
\protect\delta _{l}$. The purple vertical line represents the frequency $%
\protect\omega _{c_{1}}$ of cavity $1,$ while the blue, green, ..., and red
vertical lines represent the frequency $\protect\omega _{c_{2}}$ of cavity $2
$, the frequency $\protect\omega _{c_{3}}$ of cavity $3$,..., and the
frequency $\protect\omega _{c_{n}}$ of cavity $n$, respectively.}
\label{fig:2}
\end{figure}

Under the above assumptions and by considering the ideal cavities, the Hamiltonian of the whole system, in the
interaction picture and after making the rotating-wave approximation (RWA),
can be written as (in units of $\hbar =1$)
\begin{equation}
H_{\mathrm{I}}=g_{1}(e^{-i\delta _{1}t}\hat{a}_{1}^{+}\sigma
_{fg}^{-}+h.c.)+\sum\limits_{l=2}^{n}g_{l}(e^{-i\delta _{l}t}\hat{a}%
_{l}^{+}\sigma _{fe}^{-}+h.c.),
\end{equation}%
where $\sigma _{fg}^{-}=|g\rangle \langle f|$, $\sigma _{fe}^{-}=|e\rangle
\langle f|$, $\delta _{1}=\omega _{fg}-\omega _{c_{1}}>0,$ and $\delta
_{l}=\omega _{fe}-\omega _{c_{l}}>0.$ The detunings $\delta _{1}$ and $%
\delta _{l}$ have a relationship $\delta _{l}=\delta _{1}+\Delta _{1l},$
with $\Delta _{1l}=\omega _{c_{1}}-\omega _{c_{l}}-\omega _{eg}>0$ (Fig. 2).
Here, $\hat{a}_{1}$ ($\hat{a}_{l}$) is the photon annihilation operator of
cavity $1$ ($l$)$,$ $\omega _{c_{l}}$ is the frequency of cavity $l$ ($%
l=2,3,...,n$); while $\omega _{fg},$ $\omega _{fe},$ and $\omega _{eg}$ are
the $|f\rangle \leftrightarrow |g\rangle ,$ $|f\rangle \leftrightarrow
|e\rangle ,$ and $|e\rangle \leftrightarrow |g\rangle $ transition
frequencies of the qutrit, respectively.

Under the large-detuning conditions $\delta _{1}\gg g_{1}$ and $\delta
_{l}\gg g_{l}$, the Hamiltonian (2) becomes [77]
\begin{align}
H_{\mathrm{e}}=& -\lambda _{1}(\hat{a}_{1}^{+}\hat{a}_{1}|g\rangle \langle
g|-\hat{a}_{1}\hat{a}_{1}^{+}|f\rangle \langle f|)  \notag \\
& -\sum\limits_{l=2}^{n}\lambda _{l}(\hat{a}_{l}^{+}\hat{a}_{l}|e\rangle
\langle e|-\hat{a}_{l}\hat{a}_{l}^{+}|f\rangle \langle f|)  \notag \\
& -\sum\limits_{l=2}^{n}\lambda _{1l}(e^{i\bigtriangleup _{1l}t}\hat{a}%
_{1}^{+}\hat{a}_{l}\sigma _{eg}^{-}+h.c.)  \notag \\
& +\sum\limits_{k\neq l;k,l=2}^{n}\lambda _{kl}\left( e^{i\bigtriangleup
_{kl}t}\hat{a}_{k}^{+}\hat{a}_{l}+h.c.\right) \left( |f\rangle \langle
f|-|e\rangle \langle e|\right) ,
\end{align}%
where $\lambda _{1}=g_{1}^{2}/\delta _{1}$, $\lambda _{l}=g_{l}^{2}/\delta
_{l}$, $\lambda _{1l}=\left( g_{1}g_{l}/2\right) (1/\delta _{1}+1/\delta
_{l})$, $\lambda _{kl}=\left( g_{k}g_{l}/2\right) (1/\delta _{k}+1/\delta
_{l}),$ $\bigtriangleup _{1l}=\delta _{l}-\delta _{1}=\omega _{c_{1}}-\omega
_{c_{l}}-\omega _{eg},$ $\bigtriangleup _{kl}=\delta _{l}-\delta _{k}$ $%
=\omega _{c_{k}}-\omega _{c_{l}},$ and $\sigma _{eg}^{-}=|g\rangle \langle
e| $. In Eq.~(3), the terms in the first two lines describe the photon
number dependent stark shifts of the energy levels $|g\rangle $, $|e\rangle $
and $|f\rangle $; the terms in the third line describe the $|e\rangle $ $%
\leftrightarrow $ $|g\rangle $ coupling caused due to the cooperation of
cavities $1$ and $l$; while the terms in the last line describe the coupling
between cavities $k$ and $l.$ For $\bigtriangleup _{1l}\gg \{\lambda
_{1},\lambda _{l},\lambda _{1l},\lambda _{kl}\}$, the effective Hamiltonian $%
H_{\mathrm{e}}$ changes to [77]
\begin{align}
H_{\mathrm{e}}=-& \lambda _{1}(\hat{a}_{1}^{+}\hat{a}_{1}|g\rangle \langle
g|-\hat{a}_{1}\hat{a}_{1}^{+}|f\rangle \langle f|)  \notag \\
& -\sum\limits_{l=2}^{n}\lambda _{l}(\hat{a}_{l}^{+}\hat{a}_{l}|e\rangle
\langle e|-\hat{a}_{l}\hat{a}_{l}^{+}|f\rangle \langle f|)  \notag \\
& -\sum\limits_{l=2}^{n}\chi _{1l}(\hat{a}_{1}^{+}\hat{a}_{1}\hat{a}_{l}\hat{%
a}_{l}^{+}|g\rangle \langle g|-\hat{a}_{1}\hat{a}_{1}^{+}\hat{a}_{l}^{+}\hat{%
a}_{l}|e\rangle \langle e|)  \notag \\
& +\sum\limits_{k\neq l;k,l=2}^{n}\lambda _{kl}\left( e^{i\bigtriangleup
_{kl}t}\hat{a}_{k}^{+}\hat{a}_{l}+h.c.\right) \left( |f\rangle \langle
f|-|e\rangle \langle e|\right) ,
\end{align}%
where $\chi _{1l}=\lambda _{1l}^{2}/\Delta _{1l}$. From this equation, one
can see that each term is associated with the level $|g\rangle $, $|e\rangle
$, or $|f\rangle $. When the levels $|e\rangle $ and $|f\rangle $ are
initially not occupied, they will remain unpopulated because neither $\left\vert g\right\rangle \rightarrow \left\vert e\right\rangle $
nor $\left\vert g\right\rangle \rightarrow \left\vert f\right\rangle $ is induced under the Hamiltonian
(4). Hence, the Hamiltonian (4) reduces to
\begin{equation}
H_{\mathrm{e}}=-\lambda _{1}\hat{a}_{1}^{+}\hat{a}_{1}|g\rangle \langle
g|-\sum\limits_{l=2}^{n}\chi _{1l}\hat{a}_{1}^{+}\hat{a}_{1}\hat{a}_{l}\hat{a%
}_{l}^{+}|g\rangle \langle g|.
\end{equation}%
Note that $\left[ a_{l},a_{l}^{+}\right] =1,$ i.e., $\hat{a}_{l}\hat{a}%
_{l}^{+}=1+\hat{a}_{l}^{+}\hat{a}_{l}.$ Thus, the Hamiltonian (5) can be
rewritten as
\begin{equation}
H_{\mathrm{e}}=-\lambda _{1}\hat{n}_{1}|g\rangle \langle
g|-\sum\limits_{l=2}^{n}\chi _{1l}\hat{n}_{1}|g\rangle \langle
g|-\sum\limits_{l=2}^{n}\chi _{1l}\hat{n}_{1}\hat{n}_{l}|g\rangle \langle g|,
\end{equation}%
where $\hat{n}_{1}=\hat{a}_{1}^{+}\hat{a}_{1}$ and $\hat{n}_{l}=\hat{a}%
_{l}^{+}\hat{a}_{l}$ are the photon number operators for cavities $1$ and $l$%
, respectively.

Assume that the qutrit is initially in the ground state $\left\vert
g\right\rangle $. It will remain in this state because the Hamiltonian (6)
cannot induce any transition for the qutrit. Therefore, the Hamiltonian $H_{%
\mathrm{e}}$ reduces to
\begin{equation}
\widetilde{H}_{\mathrm{e}}=-\eta \hat{n}_{1}-\chi \sum\limits_{l=2}^{n}\hat{n%
}_{1}\hat{n}_{l},
\end{equation}%
where $\eta =\lambda _{1}+\left( n-1\right) \chi .$ Here, we have set $\chi
_{1l}=\chi $ ($l=2,3,...,n$)$.$ Note that the $\widetilde{H}_{\mathrm{e}}$
is the effective Hamiltonian governing the dynamics of the $n$ cavities ($%
1,2,...,n$).

The unitary operator $U=e^{-i\widetilde{H}_{\mathrm{e}}t}$ can be written as
\begin{equation}
U=U_{1}\left[ \otimes _{l=2}^{n}U_{1l}\right] ,
\end{equation}%
with
\begin{eqnarray}
U_{1} &=&\exp \left( i\eta \hat{n}_{1}t\right) , \\
U_{1l} &=&\exp \left( i\chi \hat{n}_{1}\hat{n}_{l}t\right) ,
\end{eqnarray}%
where $\otimes _{l=2}^{n}U_{1l}=U_{12}U_{13}...U_{1n}.$ Here, $U_{1}$ is a
unitary operator on cavity $1,$ while $U_{1l}$ is a unitary operator on
cavities $1$ and $l.$

Let us now consider $n$ MP qubits $1,$ $2,...,$and $n$, which
are hosted by cavities $1,$ $2,...,$ and $n$, respectively. The two logical
states of MP qubit $l^{\prime }$ are represented by the
vacuum state $|0\rangle $ and the single-photon state $|1\rangle $ of cavity
$l^{\prime }$ ($l^{\prime }=1,2,...,n$). Based on Eq. (10), one can easily
see that for $\chi t=\pi ,$ the unitary operation $U_{1l}$ leads to the
following state transformation
\begin{align}
U_{1l}|0_{1}0_{l}\rangle & =|0_{1}0_{l}\rangle ,  \notag \\
U_{1l}|0_{1}1_{l}\rangle & =|0_{1}1_{l}\rangle ,  \notag \\
U_{1l}|1_{1}0_{l}\rangle & =|1_{1}0_{l}\rangle ,  \notag \\
U_{1l}|1_{1}1_{l}\rangle & =-|1_{1}1_{l}\rangle ,
\end{align}%
which implies that the operator $U_{1l}$ implements a universal
controlled-phase gate on two qubits $1$ and $l.$ Eq.~(11) can be expressed
as
\begin{eqnarray}
U_{1l}|0_{1}i_{l}\rangle |g\rangle &=&|0_{1}i_{l}\rangle |g\rangle  \notag \\
U_{1l}|1_{1}i_{l}\rangle |g\rangle &=&\left( -1\right)
^{i_{l}}|1_{1}i_{l}\rangle |g\rangle ,
\end{eqnarray}%
where $i_{l}$ $\in \left\{ 0,1\right\} .$

Based on Eq.~(12), one can easily obtain the following state transformation%
\begin{eqnarray}
\otimes _{l=2}^{n}U_{1l}\left\vert 0_{1}\right\rangle \left\vert
i_{2}\right\rangle \left\vert i_{3}\right\rangle ...\left\vert
i_{n}\right\rangle &=&\left\vert 0_{1}\right\rangle \left\vert
i_{2}\right\rangle \left\vert i_{3}\right\rangle ...\left\vert
i_{n}\right\rangle ,  \notag \\
\otimes _{l=2}^{n}U_{1l}\left\vert 1_{1}\right\rangle \left\vert
i_{2}\right\rangle \left\vert i_{3}\right\rangle ...\left\vert
i_{n}\right\rangle &=&\left\vert 1_{1}\right\rangle \left( -1\right)
^{i_{2}}\left( -1\right) ^{i_{3}}...\left( -1\right) ^{i_{n}}\left\vert
i_{2}\right\rangle \left\vert i_{3}\right\rangle ...\left\vert
i_{n}\right\rangle .
\end{eqnarray}

According to (9), one can see that for $\eta t=2m\pi $ ($m$ is a positive
integer), the unitary operator $U_{1}$ leads to
\begin{equation}
U_{1}|0_{1}\rangle =|0_{1}\rangle ,\text{ }U_{1}|1_{1}\rangle =|1_{1}\rangle
.
\end{equation}%
Combining Eq.~(13) and Eq.~(14), we have
\begin{eqnarray}
U_{1}\left[ \otimes _{l=2}^{n}U_{1l}\right] \left\vert 0_{1}\right\rangle
\left\vert i_{2}\right\rangle \left\vert i_{3}\right\rangle ...\left\vert
i_{n}\right\rangle &=&\left\vert 0_{1}\right\rangle \left\vert
i_{2}\right\rangle \left\vert i_{3}\right\rangle ...\left\vert
i_{n}\right\rangle ,  \notag \\
U_{1}\left[ \otimes _{l=2}^{n}U_{1l}\right] \left\vert 1_{1}\right\rangle
\left\vert i_{2}\right\rangle \left\vert i_{3}\right\rangle ...\left\vert
i_{n}\right\rangle &=&\left\vert 1_{1}\right\rangle \left( -1\right)
^{i_{2}}\left( -1\right) ^{i_{3}}...\left( -1\right) ^{i_{n}}\left\vert
i_{2}\right\rangle \left\vert i_{3}\right\rangle ...\left\vert
i_{n}\right\rangle ,
\end{eqnarray}%
which shows that when the control qubit $1$ is in the state $\left\vert
1\right\rangle $, a phase flip (from sign $+$ to $-$) happens to the state $%
\left\vert 1\right\rangle $ of each of target qubits ($2,3,...,n$), while
nothing happens to the states of each of target qubit ($2,3,...,n$) when the
control qubit $1$ is in the state $\left\vert 0\right\rangle .$ From Eq.
(8), it can be seen that the jointed unitary operators $U_{1}\left[ \otimes
_{l=2}^{n}U_{1l}\right] $ involved in Eq.~(15) is equivalent to the unitary
operator $U.$ By comparing Eq.~(15) with Eq.~(1), one can see that a
multi-target-qubit controlled phase gate, described by Eq. (1), is realized
with $n$ MP qubits ($1,2,...,n$), after the above operation,
described by the unitary operator $U.$

We stress that the gate is realized through a single unitary operator $U,$
which was obtained by starting with the original Hamiltonian (2). In this
sense, the gate is implemented with only a single operation. In addition, it
is noted that the qutrit remains in the ground state $\left\vert
g\right\rangle $ during the gate operation. Hence, decoherence from the
qutrit is greatly suppressed.

In above, we have set $\chi _{1l}=\chi $, which turns out into
\begin{equation}
\frac{g_{1}^{2}g_{l}^{2}}{4\Delta _{1l}}(\frac{1}{\delta _{1}}+\frac{1}{%
\delta _{l}})^{2}=\chi .
\end{equation}%
In addition, we have set $\chi t=\pi $ and $\eta t=2m\pi ,$ from which we
obtain
\begin{equation}
\frac{g_{1}^{2}}{\delta _{1}}=\left( 2m-n+1\right) \chi .
\end{equation}%
Given $g_{1},$ $\delta _{1},$ $m,$ and $n,$ the value of $\chi $ can be
calculated based on Eq.~(17). In addition, given $g_{1},$ $\delta _{1},$ and
$\chi ,$ Eq.~(16) can be satisfied by varying $g_{l}$ or $\delta _{l}$ or
both. Note that the detuning $\delta _{l}$ can be adjusted by varying the
frequency of cavity $l$, and the coupling strength $g_{l}$ can be adjusted
by a prior design of the sample with appropriate capacitance or inductance
between the qutrit and cavity $l$ [13,78].

As shown above, the Hamiltonian (5) was obtained from the Hamiltonian (4)
when the levels $|e\rangle $ and $|f\rangle $ are initially not occupied.
This derivation has nothing to do with $\bigtriangleup _{kl}$. In this
sense, one can have $\bigtriangleup _{kl}\neq 0$ or $\bigtriangleup _{kl}=0$%
. Note that $\bigtriangleup _{kl}=\delta _{l}-\delta _{k}$ $=\omega
_{c_{k}}-\omega _{c_{l}}.$ Thus, the frequencies of cavities ($2,3,...,n$)
can be chosen to be different or the same. However, it is suggested that for
circuit QED, the frequencies of cavities should be different in order to
suppress the unwanted inter-cavity crosstalk.

\begin{figure}[tbp]
\begin{center}
\includegraphics[bb=180 520 440 749, width=8.5 cm, clip]{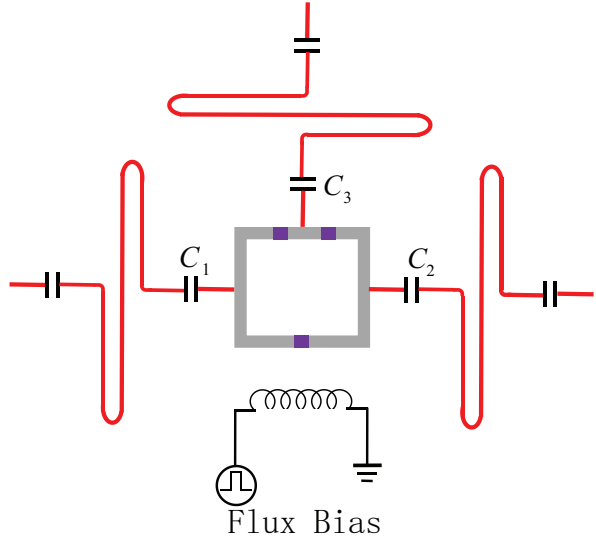} \vspace*{%
-0.08in}
\end{center}
\caption{Setup for three one-dimensional transmission line
resonators capacitively coupled to a superconducting flux qutrit.}
\label{fig:3}
\end{figure}

\section{Possible experimental implementation}

In this section, we briefly discuss the experimental feasibility of
realizing a three-qubit controlled phase gate with one MP
qubit simultaneously controlling two target MP qubits, by
considering a setup of three microwave cavities ($1,2,3$) coupled to a
superconducting flux qutrit (Fig.~3). Each cavity considered in Fig.~3 is a
one-dimensional transmission line resonator (TLR).

\begin{figure}[tbp]
\begin{center}
\includegraphics[bb=184 474 417 735, width=7.5 cm, clip]{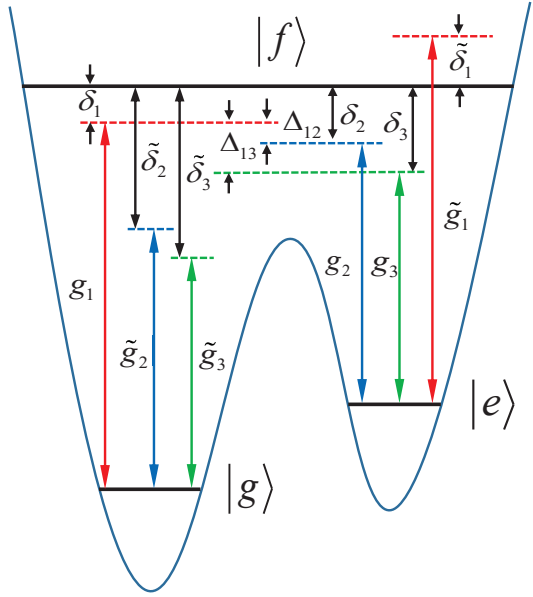} \vspace*{%
-0.08in}
\end{center}
\caption{Illustration of the unwanted coupling between cavity
$1$ and the $|e\rangle \leftrightarrow |f\rangle $ transition of the qutrit
(with coupling strength $\widetilde{g}_{1}$ and detuning $\widetilde{\protect%
\delta }_{1}$) as well as the unwanted coupling between cavity $l$ and the $%
|g\rangle \leftrightarrow |f\rangle $ transition of the qutrit (with
coupling strength $\widetilde{g}_{l}$ and detuning $\widetilde{\protect%
\delta }_{l}$) ($l=2,3$). Note that the coupling of each cavity with the $%
|g\rangle \leftrightarrow |e\rangle $ transition of the qutrit is negligible
because of the weak $|g\rangle \leftrightarrow |f\rangle $ transition.}
\label{fig:4}
\end{figure}

In reality, there exist the inter-cavity crosstalk between cavities [79],
the unwanted coupling of cavity $1$ with the $|e\rangle \leftrightarrow
|f\rangle $ transition, and the unwanted coupling of cavities $2$ and $3$
with the $|g\rangle \leftrightarrow |f\rangle $ transition of the qutrit
(Fig.~4). After taking these factors into account, the Hamiltonian (2) is
modified as
\begin{equation}
\widetilde{H}_{\mathrm{I}}=H_{\mathrm{I}}+\delta H+\varepsilon ,
\end{equation}%
with%
\begin{eqnarray}
\delta H &=&\widetilde{g}_{1}(e^{-i\widetilde{\delta }_{1}t}\hat{a}%
_{1}^{+}\sigma _{fe}^{-}+h.c.)  \notag \\
&&+\sum\limits_{l=2}^{3}\widetilde{g}_{l}(e^{-i\widetilde{\delta }_{l}t}\hat{%
a}_{l}^{+}\sigma _{fg}^{-}+h.c.),
\end{eqnarray}%
\begin{equation}
\varepsilon =\sum\limits_{k\neq l;k,l=1}^{3}g_{kl}(e^{i\widetilde{\Delta }%
_{kl}t}\hat{a}_{k}^{+}\hat{a}_{l}+h.c.).
\end{equation}%
Here, $H_{\mathrm{I}}$ is the Hamiltonian (2) for $n=3$. $\delta H$ is the
Hamiltonian, which describes the unwanted coupling between cavity $1$ and
the $|e\rangle \leftrightarrow |f\rangle $ transition with coupling strength
$\widetilde{g}_{1}$ and detuning $\widetilde{\delta }_{1}=\omega
_{fe}-\omega _{c_{1}},$ as well as the unwanted coupling between cavity $l$
and the $|g\rangle \leftrightarrow |f\rangle $ transition with coupling
strength $\widetilde{g}_{l}$ and detuning $\widetilde{\delta }_{l}=\omega
_{fg}-\omega _{c_{l}}$ ($l=2,3$) (Fig.~4). In addition, $\varepsilon $
represents the inter-cavity crosstalk, with the coupling strength $g_{kl}$
between cavities $k$ and $l$, as well as the frequency difference $%
\widetilde{\bigtriangleup }_{kl}=\omega _{c_{k}}-\omega _{c_{l}}$ of
cavities $k$ and $l$ ($k\neq l;k,l\in \left\{ 1,2,3\right\} $).

When the dissipation and dephasing are included, the dynamics of the lossy
system is determined by
\begin{align}
\frac{d\rho }{dt}=& -i[\widetilde{H}_{\mathrm{I}},\rho
]+\sum_{l=1}^{3}\kappa _{l}\mathcal{L}[a_{l}]  \notag \\
& +\gamma _{eg}\mathcal{L}[\sigma _{eg}^{-}]+\gamma _{fe}\mathcal{L}[\sigma
_{fe}^{-}]+\gamma _{fg}\mathcal{L}[\sigma _{fg}^{-}]  \notag \\
& +\sum\limits_{j=e,f}\{\gamma _{\varphi j}(\sigma _{jj}\rho \sigma
_{jj}-\sigma _{jj}\rho /2-\rho \sigma _{jj}/2)\},
\end{align}%
where $\widetilde{H}_{\mathrm{I}}$ is the above full Hamiltonian; $\sigma
_{eg}^{-}=|g\rangle \langle e|$, $\sigma _{jj}=|j\rangle \langle j|(j=e,f)$;
and $\mathcal{L}[\xi ]=\xi \rho \xi ^{\dag }-\xi ^{\dag }\xi \rho /2-\rho
\xi ^{\dag }\xi /2$, with $\xi =a_{l},\sigma _{eg}^{-},\sigma
_{fe}^{-},\sigma _{fg}^{-}$. In addition, $\kappa _{l}$ is the photon decay
rate of cavity $l$ $(l=1,2,3),$ $\gamma _{eg}$ is the energy relaxation rate
for the level $|e\rangle $ of the qutrit, $\gamma _{fe}(\gamma _{fg})$ is
the energy relaxation rate of the level $|f\rangle $ of the qutrit for the
decay path $|f\rangle \longrightarrow |e\rangle (|g\rangle )$, and $\gamma
_{\varphi j}$ is the dephasing rate of the level $|j\rangle (j=e,f)$ of the
qutrit. \newline

The fidelity of the operation is given by
\begin{equation}
\mathcal{F}=\sqrt{\langle \psi _{\mathrm{id}}|\rho |\psi _{\mathrm{id}%
}\rangle },
\end{equation}%
where $|\psi _{\mathrm{id}}\rangle $ is the output state of an ideal system
without dissipation, dephasing and crosstalk; while $\rho $ is the final
practical density operator of the system when the operation is performed in
a realistic situation. For simplicity, we consider the three qubits are
initially in the following state%
\begin{eqnarray}
|\psi _{\mathrm{in}}\rangle &=&\frac{1}{2\sqrt{2}}\left( \left\vert
000\right\rangle +\left\vert 001\right\rangle +\left\vert 010\right\rangle
+\left\vert 011\right\rangle \right.  \notag \\
&&\left. +\left\vert 100\right\rangle +\left\vert 101\right\rangle
+\left\vert 110\right\rangle +\left\vert 111\right\rangle \right) .
\end{eqnarray}
Thus, the ideal output state of the whole system is
\begin{eqnarray}
|\psi _{\mathrm{id}}\rangle &=&\frac{1}{2\sqrt{2}}\left( \left\vert
000\right\rangle +\left\vert 001\right\rangle +\left\vert 010\right\rangle
+\left\vert 011\right\rangle \right.  \notag \\
&&\left. +\left\vert 100\right\rangle -\left\vert 101\right\rangle
-\left\vert 110\right\rangle +\left\vert 111\right\rangle \right) \otimes
\left\vert g\right\rangle .
\end{eqnarray}

\begin{figure}[tbp]
\begin{center}
\includegraphics[bb=75 483 477 754, width=12.5 cm, clip]{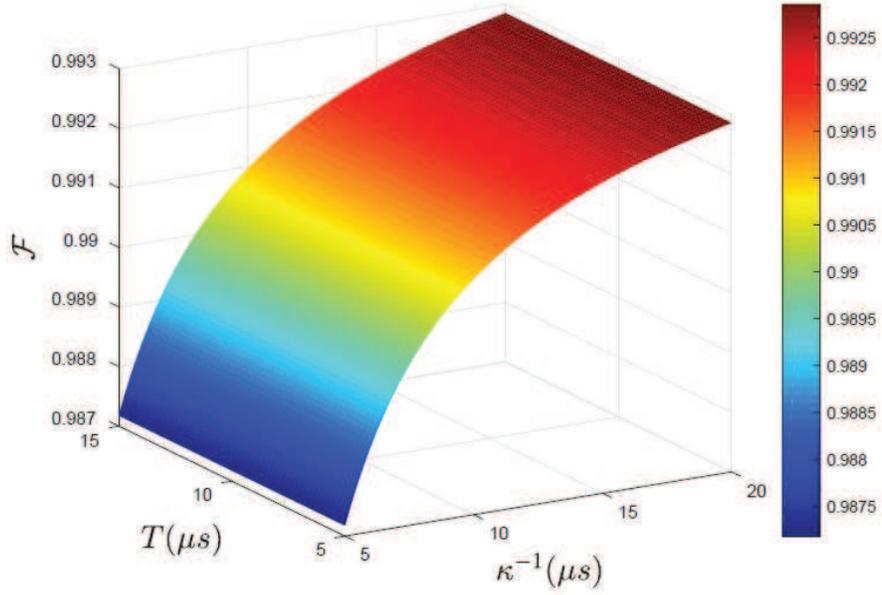} \vspace*{%
-0.08in}
\end{center}
\caption{Fidelity versus $T$ and $\protect\kappa ^{-1}$. The
parameters used in the numerical simulation are referred to the text.}
\label{fig:4}
\end{figure}

For a flux qutrit, the typical transition frequency between neighboring
levels can be made as 1 to 20 GHz. As an example, we consider $\omega
_{eg}/2\pi =5.0$ GHz, $\omega _{fe}/2\pi =7.5$ GHz, and $\omega _{fg}/2\pi
=12.5$ GHz. By choosing $\delta _{1}/2\pi =1.5$ GHz, $\delta _{2}/2\pi =1.51$
GHz, and $\delta _{3}/2\pi =1.53$ GHz, we have $\Delta _{12}/2\pi =10$ MHz, $%
\Delta _{13}/2\pi =30$ MHz, $\omega _{c_{1}}/2\pi =11$ GHz, $\omega
_{c_{2}}/2\pi =5.99$ GHz, and $\omega _{c_{3}}/2\pi =5.97$ GHz, for which we
have $\widetilde{\triangle }_{12}/2\pi =5.01$ GHz, $\widetilde{\triangle }%
_{23}/2\pi =0.02$ GHz, and $\widetilde{\triangle }_{13}/2\pi =5.03$ GHz.
With the transition frequencies of the qutrit and the frequencies of the
cavities given here, we have $\widetilde{\delta }_{1}/2\pi =-3.5$ GHz, $%
\widetilde{\delta }_{2}/2\pi =6.51$ GHz, and $\widetilde{\delta }_{3}/2\pi
=6.53$ GHz. Other parameters used in the numerical simulation are: (i) $%
\gamma _{eg}^{-1}=5T$ $\mu $s, $\gamma _{fe}^{-1}=2T$ $\mu $s, $\gamma
_{fg}^{-1}=T$ $\mu $s, (ii) $\gamma _{\phi e}^{-1}=\gamma _{\phi f}^{-1}=T$ $%
\mu $s, and (iii) $g_{1}/2\pi =150$ MHz. According to Eqs.~(16) and (17),
one can calculate the $g_{2}$ and $g_{3}$, which are $g_{2}/2\pi \sim 86.89$
MHz and $g_{3}/2\pi \sim 151.49$ MHz. For a flux qutrit, one has $\widetilde{%
g}_{1}\sim g_{1},$ $\widetilde{g}_{2}\sim g_{2},$ and $\widetilde{g}_{3}\sim
g_{3}$. Note that the coupling constants chosen here are readily available
because a coupling constant $\sim 2\pi \times 636$ MHz has been reported for
a flux device coupled to a one-dimensional transmission line resonator [40].
We set $g_{kl}=0.01g_{\max }$, where $g_{\max }=\max
\{g_{1},g_{2},g_{3}\}\sim 2\pi \times 151.49$ MHz, which can be achieved in
experiments [56,69]. In addition, assume $\kappa _{1}=\kappa _{2}=\kappa
_{3}=\kappa $ for simplicity.

By solving the master equation (21), we numerically calculate the fidelity
versus $T$ and $\kappa ^{-1}$, as depicted in Fig.~5. From Fig. 5, one can
see that when $T\geqslant 5$ $\mu $s and $\kappa ^{-1}\geqslant 10$ $\mu $s,
fidelity exceeds $0.9909,$ which implies that a high fidelity can be
obtained for the gate being performed in a realistic situation.

\begin{figure}[tbp]
\begin{center}
\includegraphics[bb=0 0 160 127, width=8.5 cm, clip]{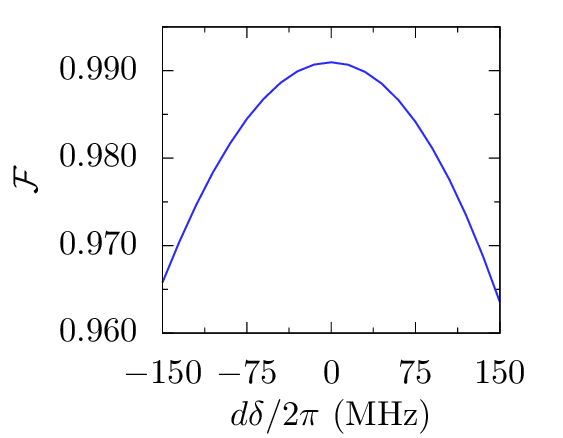} \vspace*{%
-0.08in}
\end{center}
\caption{Fidelity versus $d\delta $. Here, $d\delta $ is the
detuning error, which applies to each of detunings $\delta _{1},\delta _{2},$ and $%
\delta _{3}.$ The figure is plotted for $T=5$ $\mu $s and $\kappa ^{-1}=10$ $\mu $s. Other
parameters used in the numerical simulation are the same as those used in Fig. 5. }
\label{fig:6}
\end{figure}

To investigate the effect of the detuning errors on the fidelity, we
consider a small deviation $d\delta $ for $\delta _{1},\delta _{2},$ and $%
\delta _{3}.$ Thus, we modify  $\delta _{1},\delta _{2},$ and $\delta _{3}$
as $\delta _{1}+d\delta ,\delta _{2}+d\delta ,$ and $\delta _{3}+d\delta $.
With this modification, we numerically calculate the fidelity for $T=5$ $\mu
$s and $\kappa ^{-1}=10$ $\mu $s and plot Fig. 6 showing the fidelity
versus $d\delta .$ From Fig. 6, one can see that the fidelity can reach 0.98
or greater for $-75$ MHz $\leq d\delta /2\pi $ $\leq 75$ MHz.

The gate operational time is estimated as $\sim 66.7$ ns for the parameters
chosen above, which is much shorter than the decoherence times of the qutrit
($5$ $\mu $s $-$ $75$ $\mu $s) and the cavity decay times ($5$ $\mu $s $-$ $%
20$ $\mu $s) considered in Fig. 5. Here, we consider a rather conservative
case for decoherence time of the flux qutrit because decoherence time 70 $%
\mu $s to 1 ms has been experimentally reported for a superconducting flux
device [32,36,38]. For the cavity frequencies given above and $\kappa
^{-1}=10$ $\mu $s, one has $Q_{1}\sim 6.9\times 10^{5}$ for cavity $1$, $%
Q_{2}\sim 3.76\times 10^{5}$ for cavity $2$, and $Q_{3}\sim 3.75\times 10^{5}
$ for cavity $3$, which are available because TLRs with a (loaded) quality
factor $Q\sim 10^{6}$ have been experimentally demonstrated [42,43]. The
analysis here implies that high-fidelity realization of a quantum controlled
phase gate with one MP qubit simultaneously controlling two
target MP qubits is feasible with the present circuit QED
technology.

\begin{figure}[tbp]
\begin{center}
\includegraphics[bb=365 190 592 419, width=8.5 cm, clip]{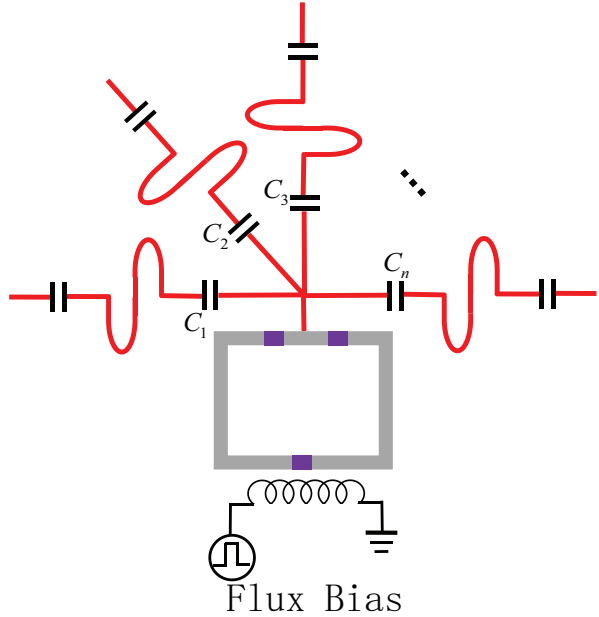} \vspace*{%
-0.08in}
\end{center}
\caption{Schematic diagram for $n$ cavities coupled by a superconducting flux qutrit.
Each cavity here is a one-dimensional transmission line resonator, which is coupled to the qutrit via a capacitor.}
\label{fig:7}
\end{figure}

In above, we have provided the specific implementation of the three qubits case.
For the gate with more than three qubits, the extension is straightforward.
From Fig. 7, one can see that each of the multiple cavities can in principle be coupled to
a single superconducting flux qutrit via a capacitor. However, it should be pointed out that in the solid-state setup
scaling up to many cavities coupled to one qutrit will introduce new challenges.
For instance, the cavity crosstalk may become worse as the number of cavities increases,
which will decrease the operation fidelity.

\section{Conclusion}
We have presented a one-step approach to realize an $n$-qubit controlled
phase gate with one microwave photonic qubit simultaneously controlling $n-1$
target microwave photonic qubits, based on circuit QED. As shown above, this
proposal has the following advantages: (i) During the gate operation, the
qutrit remains in the ground state; thus decoherence from the qutrit is
greatly suppressed; (ii) Because only one-step operation is needed and
neither classical pulse nor measurement is required, the gate implementation
is simple; (iii) The gate operation time is independent of the number of
the qubits; and (iv) This proposal is quite general and can be applied to realize the proposed gate with a wide range of
physical systems, such as multiple microwave or optical cavities coupled to
a single $\Lambda$-type three-level natural or artificial atom. Furthermore, our numerical simulations demonstrate that
high-fidelity implementation of a three-qubit controlled phase gate with one
microwave photonic qubit simultaneously controlling two target microwave
photonic qubits is feasible with present circuit QED technology. We hope that this work will
stimulate experimental activities in the near future.

\section*{Acknowledgment}
This work was supported in part by the NKRDP of China (Grant No.
2016YFA0301802) and the National Natural Science Foundation of China under
Grant Nos. [11074062, 11374083,11774076]. This work was also supported by
the Hangzhou-City grant for Quantum Information and Quantum Optics
Innovation Research Team.


\begin{thebibliography}{99}

\newcommand{\enquote}[1]{``#1''}

\bibitem{s1} L. M. Duan, B. Wang, and H. J. Kimble, \enquote{Robust quantum gates on
neutral atoms with cavity-assisted photon-scattering}, Phys. Rev. A \textbf{72%
}, 032333 (2005).

\bibitem{s2} X. Wang, A. S\o ensen, and K. M\o meret, \enquote{Multibit Gates for
Quantum Computing}, Phys. Rev. Lett. \textbf{86}, 3907 (2001).

\bibitem{s3} C. P. Yang and S. Han, \enquote{n-qubit-controlled phase gate with
superconducting quantum-interference devices coupled to a resonator}, Phys.
Rev. A \textbf{72}, 032311 (2005).

\bibitem{s4} X. Zou, Y. Dong, and G. C. Guo, \enquote{Implementing a conditional z
gate by a combination of resonant interaction and quantum interference},
Phys. Rev. A \textbf{74}, 032325 (2006).

\bibitem{s5} C. P. Yang and S. Han, \enquote{Realization of an n-qubit controlled-U
gate with superconducting quantum interference devices or atoms in cavity
QED}, Phys. Rev. A \textbf{73}, 032317 (2006).

\bibitem{s6} T. Monz, K. Kim, W. H\"{a}nsel, M. Riebe, A. S.Villar, P.
Schindler, M. Chwalla, M. Hennrich, and R. Blatt, \enquote{Realization of the Quantum
Toffoli Gate with Trapped Ions}, Phys. Rev. Lett. \textbf{102}, 040501 (2009).

\bibitem{s7} W. L. Yang, Z. Q. Yin, Z. Y. Xu, M. Feng, and J. F. Du, \enquote{One
step implementation of multi-qubit conditional phase gating with
nitrogen-vacancy centers coupled to a high-Q silicamicro sphere cavity},
Appl. Phys. Lett. \textbf{96}, 241113 (2010).

\bibitem{s8} S. B. Zheng, \enquote{Implementation of Toffoli gates with a single
asymmetric Heisenberg XY interaction}, Phys. Rev. A \textbf{87}, 042318
(2013).

\bibitem{s9} H. R. Wei and F. G. Deng, \enquote{Universal quantum gates for hybrid
systems assisted by quantum dots inside double-sided optical microcavities},
Phys. Rev. A \textbf{87}, 022305 (2013).

\bibitem{s10} H. W. Wei and F. G. Deng, \enquote{Scalable quantum computing based on
stationary spin qubits in coupled quantum dots inside double-sided optical
microcavities}, Sci. Rep. \textbf{4}, 7551 (2014).

\bibitem{s11} M. Hua, M. J. Tao, and F. G. Deng, \enquote{Universal quantum gates on
microwave photons assisted by circuit quantum electrodynamics}, Phys. Rev. A
\textbf{90}, 012328 (2014).

\bibitem{s12} M. Hua, M. J. Tao, and F. G. Deng, \enquote{Fast universal quantum
gates on microwave photons with all-resonance operations in circuit QED},
Sci. Rep. \textbf{5}, 9274 (2015).

\bibitem{s13} B. Ye, Z. F. Zheng, and C. P. Yang, \enquote{Multiplex-controlled phase
gate with qubits distributed in a multicavity system}, Phys. Rev. A \textbf{97%
}, 062336 (2018).

\bibitem{s14} Q. Wei, X. Wang, A. Miranowicz, Z. Zhong, and F. Nori, \enquote{Heralded quantum controlled-PHASE gates with dissipative dynamics
in macroscopically distant resonators}, Phys. Rev. A \textbf{96}, 012315 (2017).

\bibitem{s15} M. \v{S}a\v{s}ura and V. Buzek, \enquote{Multiparticle entanglement
with quantum logic networks: Application to cold trapped ions}, Phys. Rev. A
\textbf{64}, 012305 (2001).

\bibitem{s16} F. Gaitan, \textit{Quantum Error Correction and Fault Tolerant Quantum
Computing} (CRC Press, USA, 2008).

\bibitem{s17} T. Beth and M. R\"{o}tteler, \textit{Quantum Information}
(Springer,Berlin, 2001), Vol. 173, Ch. 4, p. 96.

\bibitem{s18} S. L. Braunstein, V. Bu\v{z}ek, and M. Hillery, \enquote{Quantum
network for symmetric and asymmetric cloning in arbitrary dimension and
continuous limit}, Phys. Rev. A \textbf{63}, 052313 (2001).

\bibitem{s19} C. P. Yang, Y. X. Liu, and F. Nori, \enquote{Phase gate of one qubit
simultaneously controlling $n$ qubits in a cavity}, Phys. Rev. A \textbf{81},
062323 (2010).

\bibitem{s20} C. P. Yang, S. B. Zheng, and F. Nori, \enquote{Multiqubit tunable phase
gate of one qubit simultaneously controlling $n$ qubits in a cavity}, Phys.
Rev. A \textbf{82}, 062326 (2010).

\bibitem{s21} H. F. Wang, A. D. Zhu, and S. Zhang, \enquote{One-step implementation
of multiqubit phase gate with one control qubit and multiple target qubits
in coupled cavities}, Opt. Lett. \textbf{39}, 1489-1492 (2014).

\bibitem{s22} C. P. Yang, Q. P. Su, F. Y. Zhang, and S. B. Zheng,
\enquote{Single-step implementation of a multiple-target-qubit controlled phase gate
without need of classical pulses}, Opt. Lett. \textbf{39}, 3312-3315 (2014).

\bibitem{s23} T. Liu, X. Z. Cao, Q. P. Su, S. J. Xiong, and C. P. Yang,
\enquote{Multi-target-qubit unconventional geometric phase gate in a multicavity
system}, Sci. Rep. \textbf{6}, 21562 (2016).

\bibitem{s24} J. Clarke and F. K. Wilhelm, \enquote{Superconducting quantum bits},
Nature \textbf{453}, 1031-1042 (2008).

\bibitem{s25} I. Buluta, S. Ashhab, and F. Nori, \enquote{Natural and artificial
atoms for quantum computation}, Rep. Prog. Phys. \textbf{74}, 104401-104416 (2011).

\bibitem{s26} J. Q. You and F. Nori, \enquote{Atomic physics and quantum optics using
superconducting circuits}, Nature (London) \textbf{474}, 589-597 (2011).

\bibitem{s27} Z. L. Xiang, S. Ashhab, J. Q. You, and F. Nori, \enquote{Hybrid quantum
circuits: Superconducting circuits interacting with other quantum systems},
Rev. Mod. Phys. \textbf{85}, 623-653 (2013).

\bibitem{s28} J. Q. You and F. Nori, \enquote{Superconducting circuits and quantum information}, Phys. Today \textbf{58}, 42-47 (2005).

\bibitem{s29} X. Gu, A. F. Kockum, A. Miranowicz, Y. X. Liu, and F. Nori, \enquote{Microwave photonics with superconducting quantum circuits}, Phys. Rep. \textbf{718}, 1-102 (2017).

\bibitem{s30} J. Bylander, S. Gustavsson, F. Yan, F. Yoshihara, K. Harrabi,
G. F. David, G. Cory, Y. Nakamura, J. S. Tsai, and W. D. Oliver, \enquote{Noise
spectroscopy through dynamical decoupling with a superconducting flux qubit},
Nat. Phys. \textbf{7}, 565-570 (2011).

\bibitem{s31} H. Paik, D. I. Schuster, L. S. Bishop, G. Kirchmair, G.
Catelani, A. P. Sears, B. R. Johnson, M. J. Reagor, L. Frunzio, L. I.
Glazman, S. M. Girvin, M. H. Devoret, and R. J. Schoelkopf, \enquote{Observation of high coherence in josephson junction qubits
measured in a three-dimensional circuit QED architecture}, Phys. Rev. Lett.
\textbf{107}, 240501 (2011).

\bibitem{s32} C. Rigetti, S. Poletto, J. M. Gambetta, B. L. T. Plourde, J.
M. Chow, A. D. Corcoles, J. A. Smolin, S. T. Merkel, J. R. Rozen, G. A.
Keefe, M. B. Rothwell, M. B. Ketchen, and M. Steffen, \enquote{Superconducting qubit in waveguide cavity with coherence time
approaching 0.1 ms}, Phys. Rev. B \textbf{86}, 100506(R) (2012).

\bibitem{s33} R. Barends, J. Kelly, A. Megrant, D. Sank, E. Jeffrey, Y.
Chen, Y. Yin, B. Chiaro, J. Mutus, C. Neill, P. O'Malley, P. Roushan, J. Wenner, T. C. White, A. N. Cleland,
and J. M. Martinis, \enquote{Coherent Josephson
qubit suitable for scalable quantum integrated circuits}, Phys. Rev. Lett.
\textbf{111}, 080502 (2013).

\bibitem{s34} Y. Chen, C. Neill, P. Roushan, N. Leung, M. Fang, R. Barends,
J. Kelly, B. Campbell, Z. Chen, B. Chiaro, A. Dunsworth, E. Jeffrey, A. Megrant, J. Y. Mutus,
P. J. J. O'Malley, C. M. Quintana, D. Sank, A. Vainsencher, J. Wenner, T. C. White,
Michael R. Geller, A. N. Cleland, and J. M. Martinis, \enquote{Qubit architecture
with high coherence and fast tunable coupling}, Phys. Rev. Lett. \textbf{113}, 220502 (2014).

\bibitem{s35} M. Stern, G. Catelani, Y. Kubo, C. Grezes, A. Bienfait, D.
Vion, D. Esteve, and P. Bertet, \enquote{Flux qubits with long coherence times for
hybrid quantum circuits}, Phys. Rev. Lett. \textbf{113}, 123601 (2014).

\bibitem{s36} I. M. Pop, K. Geerlings, G. Catelani, R. J. Schoelkopf, L. I.
Glazman, and M. H. Devoret, \enquote{Coherent suppression of electromagnetic
dissipation due to superconducting quasiparticles}, Nature (London) \textbf{%
508}, 369-372 (2014).

\bibitem{s37} M. J. Peterer, S. J. Bader, X. Jin, F. Yan, A. Kamal, T. J.
Gudmundsen, P. J. Leek, T. P. Orlando, W. D. Oliver, and S. Gustavsson,
\enquote{Coherence and decay of higher energy levels of a superconducting transmon
qubit}, Phys. Rev. Lett. \textbf{114}, 010501 (2015).

\bibitem{s38} F. Yan, S. Gustavsson, A. Kamal, J. Birenbaum, A. P. Sears, D.
Hover, T. J. Gudmundsen, J. L. Yoder, T. P. Orlando, J. Clarke, A. J. Kerman, and W. D. Oliver,
\enquote{The Flux Qubit Revisited to Enhance Coherence and Reproducibility}, Nat.
Commun. \textbf{7}, 12964 (2016); J. Q. You, X. Hu, S. Ashhab, and F. Nori, \enquote{Low-decoherence flux qubit},
Phys. Rev. B \textbf{75}, 140515(R) (2007).

\bibitem{s39} A. Wallraff, D. I. Schuster, A. Blais, L. Frunzio, R. S.
Huang, J. Majer, S. Kumar, S. M. Girvin, and R. J. Schoelkopf, \enquote{Strong
coupling of a single photon to a superconducting qubit using circuit quantum
electrodynamics}, Nature (London) \textbf{431}, 162-167 (2004).

\bibitem{s40} T. Niemczyk, F. Deppe, H. Huebl, E. P. Menzel, F. Hocke, M. J.
Schwarz, J. J. Garcia Ripoll, D. Zueco, T. H\"{u}mmer, E. Solano, A. Marx, and R. Gross,
\enquote{Circuit quantum electrodynamics in the ultrastrong-coupling regime}, Nat.
Phys. \textbf{6}, 772-776 (2010).

\bibitem{s41} A. F. Kockum, A. Miranowicz, S. D. Liberato, S. Savasta, and F. Nori, \enquote{Ultrastrong coupling between light and matter}, arXiv:1807.11636.

\bibitem{s42} W. Chen, D. A. Bennett, V. Patel, and J. E. Lukens, \enquote{Substrate
and process dependent losses in superconducting thin film resonators},
Supercond. Sci. Technol. \textbf{21}, 075013 (2008).

\bibitem{s43} P. J. Leek, M. Baur, J. M. Fink, R. Bianchetti, L. Steffen, S.
Filipp, and A. Wallraff, \enquote{Cavity quantum electrodynamics with separate photon
storage and qubit readout modes}, Phys. Rev. Lett. \textbf{104}, 100504
(2010).

\bibitem{s44} M. Reagor, W. Pfaff, C. Axline, R. W. Heeres, N. Ofek, K.
Sliwa, E. Holland, C. Wang, J. Blumoff, K. Chou, M. J. Hatridge, L. Frunzio, M. H. Devoret,
L. Jiang, and R. J. Schoelkopf, \enquote{A quantum memory
with near-millisecond coherence in circuit QED}, Phys. Rev. B \textbf{94},
014506 (2016).

\bibitem{s45} C. P. Yang, S. I. Chu, and S. Han, \enquote{Possible realization of
entanglement, logical gates, and quantum-information transfer with
superconducting-quantum-interference-device qubits in cavity QED}, Phys. Rev.
A \textbf{67}, 042311 (2003).

\bibitem{s46} J. Q. You and F. Nori, \enquote{Quantum information processing with
superconducting qubits in a microwave field}, Phys. Rev. B \textbf{68},
064509 (2003).

\bibitem{s47} A. Blais, R. S. Huang, A. Wallraff, S. M. Girvin, and R. J.
Schoelkopf, \enquote{Cavity quantum electrodynamics for superconducting electrical
circuits:An architecture for quantum computation}, Phys. Rev. A \textbf{69},
062320 (2004).

\bibitem{s48} M. Hofheinz, H. Wang, M. Ansmann, R. C. Bialczak, E. Lucero,
M. Neeley, A. D. O'Connell, D. Sank, J. Wenner, J. M. Martinis, and A. N.
Cleland, \enquote{Synthesizing arbitrary quantum states in a superconducting
resonator}, Nature (London) \textbf{459}, 546-549 (2009).

\bibitem{s49} H. Wang, M. Hofheinz, J. Wenner, M. Ansmann, R. C. Bialczak,
M. Lenander, E. Lucero, M. Neeley, A. D. O'Connell, D. Sank, M. Weides, A.
N. Cleland, and J. M. Martinis, \enquote{Improving the Coherence Time of
Superconducting Coplanar Resonators}, Appl. Phys. Lett. \textbf{95}, 233508
(2009).

\bibitem{s50} M. H. Devoret and R. J. Schoelkopf, \enquote{Superconducting circuits
for quantum information: an outlook}, Science \textbf{339}, 1169-1174 (2013).

\bibitem{s51} Y. X. Liu, L. F. Wei, and F. Nori, \enquote{Generation of nonclassical
photon states using a supercon ducting qubit in a microcavity}, Europhys.
Lett. \textbf{67}, 941-947 (2004).

\bibitem{s52} M. Hofheinz, E. M. Weig, M. Ansmann, R. C. Bialczak, E.
Lucero, M. Neeley, A. D. O'Connell, H. Wang, J. M. Martinis, and A. N.
Cleland, \enquote{Generation of fock states in a superconducting quantum circuit},
Nature (London) \textbf{454}, 310-314 (2008).

\bibitem{s53} F. W. Strauch, K. Jacobs, and R. W. Simmonds, \enquote{Arbitrary control
of entanglement between two superconducting resonators}, Phys. Rev. Lett.
\textbf{105}, 050501 (2010).

\bibitem{s54} Q. P. Su, C. P. Yang, and S. B. Zheng, \enquote{Fast and simple scheme
for generating NOON states of photons in circuit QED}, Sci. Rep. \textbf{4},
3898 (2014).

\bibitem{s55} M. Hua, M. J. Tao, and F. G. Deng. \enquote{Universal quantum gates on
microwave photons assisted by circuit quantum electrodynamics}, Phys. Rev. A
\textbf{90}, 18824 (2014).

\bibitem{s56} S. J. Xiong, Z. Sun, J. M. Liu, T. Liu, and C. P. Yang,
\enquote{Efficient scheme for generation of photonic NOON states in circuit QED}, Opt.
Lett. \textbf{40}, 2221-2224 (2015).

\bibitem{s57} M. Hua, M. J. Tao, and F. G. Deng, \enquote{Quantum state transfer and
controlled-phase gate on one-dimensional superconducting resonators assisted
by a quantum bus}, Sci. Rep. \textbf{6}, 22037 (2016).

\bibitem{s58} A. N. Korotkov, \enquote{Flying microwave qubits with nearly perfect
transfer efficiency}, Phys. Rev. B \textbf{84}, 014510 (2011).

\bibitem{s59} E. A. Sete, E. Mlinar, and A. N. Korotkov, \enquote{Robust quantum
state transfer using tunable couplers}, Phys. Rev. B \textbf{91}, 144509
(2015).

\bibitem{s60} H. Wang, M. Mariantoni, R. C. Bialczak, M. Lenander, E.
Lucero, M. Neeley, A. D. O'Connell, D. Sank, M. Weides, J. Wenner, T. Yamamoto, Y. Yin, J. Zhao, J. M. Martinis,
and A. N. Cleland, \enquote{Deterministic entanglement of photons in two superconducting microwave
resonators}, Phys. Rev. Lett. \textbf{106}, 060401 (2011).

\bibitem{s61} S. J. Srinivasan, N. M. Sundaresan, D. Sadri, Y. Liu, J. M.
Gambetta, T. Yu, S. M. Girvin, and A. A. Houck, \enquote{Time-reversal symmetrization
of spontaneous emission for quantum state transfer}, Phys. Rev. A \textbf{89}%
, 033857 (2014).

\bibitem{s62} J. Wenner, Y. Yin, Y. Chen, R. Barends, B. Chiaro, E. Jeffrey, J. Kelly, A. Megrant, J. Y. Mutus, C. Neill, P. J. J. O'Malley, P. Roushan, D. Sank, A. Vainsencher, T. C. White, A. N. Korotkov, A. N. Cleland, and J. M. Martinis, \enquote{Catching Time-Reversed Microwave
    Coherent State Photons with $99.4\%$ Absorption Efficiency}, Phys. Rev. Lett. \textbf{%
112}, 210501 (2014).

\bibitem{s63} C. P. Yang, Q. P. Su, and S. Y. Han, \enquote{Generation of
Greenberger-Horne-Zeilinger entangled states of photons in multiple cavities
via a superconducting qutrit or an atom through resonant interaction}, Phys.
Rev. A \textbf{86}, 022329 (2012).

\bibitem{s64} C. P. Yang, Q. P. Su, S. B. Zheng, and S. Han, \enquote{Generating
entanglement between microwave photons and qubits in multiple cavities
coupled by a superconducting qutrit}, Phys. Rev. A \textbf{87}, 022320 (2013).

\bibitem{s65} S. E. Nigg, \enquote{Deterministic hadamard gate for microwave
cat-state qubits in circuit QED}, Phys. Rev. A \textbf{89}, 022340 (2014).

\bibitem{s66} R. W. Heeres, P. Reinhold, N. Ofek, L. Frunzio, L. Jiang, M.
H. Devoret, and R. J. Schoelkopf, \enquote{Implementing a universal gate set on a
logical qubit encoded in an oscillator}, arXiv:1608.02430 (2016).

\bibitem{s67} C. P. Yang, Q. P. Su, S. B. Zheng, F. Nori, and S. Han,
\enquote{Entangling two oscillators with arbitrary asymmetric initial states}, Phys.
Rev. A \textbf{95}, 052341 (2017).

\bibitem{s68} C. Wang, Y. Y. Gao, P. Reinhold, R. W. Heeres, N. Ofek, K.
Chou, C. Axline, M. Reagor, J. Blumoff, K. M. Sliwa, L. Frunzio, S. M. Girvin, L. Jiang,
M. Mirrahimi, M. H. Devoret, and R. J. Schoelkopf, \enquote{A Schrodinger Cat Living in Two Boxes},
Science \textbf{352}, 1087-1091 (2016).

\bibitem{s69} Y. Zhang, X. Zhao, Z. F. Zheng, L. Yu, Q. P. Su, and C. P.
Yang, \enquote{Universal controlled-phase gate with cat-state qubits in circuit QED},
Phys. Rev. A \textbf{96}, 052317 (2017).

\bibitem{s70} H. F. Wang, A. D. Zhu, S. Zhang, and K. H. Yeon, \enquote{Deterministic CNOT gate and entanglement swapping for
photonic qubits using a quantum-dot spin in a double-sided optical microcavity},
Phys. Lett. A \textbf{377}, 2870 (2013).

\bibitem{s71} C. H. Bai, D. Y. Wang, S. Hu, W. X. Cui, X. X. Jiang, and H. F. Wang, \enquote{Scheme for implementing multitarget qubit controlled-NOT gate of photons and controlled-phase gate of electron spins via quantum dot-microcavity coupled system},
Quantum. Inf. Process \textbf{15}, 1485-1498 (2016).

\bibitem{s72} J. R. Johansson, N. Lambert, I. Mahboob, H. Yamaguchi, and F. Nori, \enquote{Entangled-state generation and Bell inequality violations in nanomechanical resonators}, Phys. Rev. B \textbf{90}, 174307 (2014).

\bibitem{s73} P. J. Leek, S. Filipp, P. Maurer, M. Baur, R. Bianchetti, J.
M. Fink, M. G\"{o}ppl, L. Steffen, and A. Wallraff, \enquote{Using sideband
transitions for two-qubit operations in superconducting circuits}, Phys. Rev.
B \textbf{79}, 180511 (2009).

\bibitem{s74} M. Neeley, M. Ansmann, R. C. Bialczak, M. Hofheinz, N. Katz,
E. Lucero, A. O'Connell, H. Wang, A. N. Cleland, and J. M. Martinis, \enquote{Process
tomography of quantum memory in a Josephson-phase qubit coupled to a
two-level state}, Nat. Phys. \textbf{4}, 523-526 (2008).

\bibitem{s75} M. Sandberg, C. M. Wilson, F. Persson, T. Bauch, G. Johansson,
V. Shumeiko, T. Duty, and P. Delsing, \enquote{Tuning the field in a
microwave resonator faster than the photon life time}, Appl. Phys. Lett. \textbf{92}%
, 203501 (2008).

\bibitem{s76} Z. L. Wang, Y. P. Zhong, L. J. He, H. Wang, J. M. Martinis, A.
N. Cleland, and Q. W. Xie, \enquote{Quantum state characterization of a fast tunable
superconducting resonator}, Appl. Phys. Lett. \textbf{102}, 163503 (2013).

\bibitem{s77} D. F. James and J. Jerke, \enquote{Effective Hamiltonian theory and its
applications in quantum information}, Can. J. Phys. \textbf{85}, 625-632 (2007).

\bibitem{s78} Qi-Ping Su, H. H. Zhu, L. Yu, Y. Zhang, S. J. Xiong, J. M.
Liu, and C. P. Yang, \enquote{Generating double NOON states of photons in circuit
QED}, Phys. Rev. A \textbf{95}, 022339 (2017).

\bibitem{s79} C. P. Yang, Q. P. Su, S. B. Zheng, and F. Nori,
\enquote{Crosstalk-insensitive method for simultaneously coupling multiple pairs of
resonators}, Phys. Rev. A \textbf{93}, 042307 (2016).

\end{thebibliography}
\end{document}